\documentclass[letterpaper, 10 pt, conference]{ieeeconf}  % Comment this line out if you need a4paper

\IEEEoverridecommandlockouts                              
\usepackage[utf8]{inputenc}
\usepackage{color}
\usepackage{array}

\usepackage{float}
\usepackage{amsmath}
\usepackage{amssymb}
\usepackage{graphicx}
\usepackage{longtable}
\usepackage{multirow}

\usepackage{verbatim}
\usepackage{amsmath}
\usepackage{amssymb}
\usepackage{graphicx}

\usepackage[unicode=true,
bookmarks=false,
breaklinks=false,pdfborder={0 0 1},colorlinks=false]
{hyperref}
\hypersetup{
	colorlinks,bookmarksopen=red,bookmarksnumbered=red,citecolor=blue,urlcolor=red}
\usepackage{cite}

\floatstyle{ruled}
\newfloat{algorithm}{tbp}{loa}
\providecommand{\algorithmname}{Algorithm}
\floatname{algorithm}{\protect\algorithmname}

\newtheorem{defn}{Definition}

\newtheorem{lem}{Lemma}

\newtheorem{thm}{Theorem}

\newtheorem{assum}{Assumption}
\usepackage{cite}

\title{\LARGE \bf Stochastic Observer for SLAM on the Lie Group }

\author{Marium Tawhid, Ajay Singh Ludher, and Hashim A. Hashim\\% <-this % stops a space
		Software Engineering\\
	Department of Engineering and Applied Science\\
	Thompson Rivers University,	Kamloops, British Columbia, Canada, V2C-0C8\\
	tawhidm16@mytru.ca, ludhera17@mytru.ca, and hhashim@tru.ca
\thanks{This work was supported in part by Thompson Rivers University Internal research fund, RGS-2020/21 IRF, \# 102315.}}% <-this % stops a space

\begin{document}
\bstctlcite{IEEEexample:BSTcontrol}

\maketitle
\thispagestyle{empty}
\pagestyle{empty}

%%%%%%%%%%%%%%%%%%%%%%%%%%%%%%%%%%%%%%%%%%%%%%%%%%%%%%%%%%%%%%%%%%%%%%%%%%%%%%%%
\begin{abstract}
A robust nonlinear stochastic observer for simultaneous localization
and mapping (SLAM) is proposed using the available uncertain measurements
of angular velocity, translational velocity, and features. The proposed
observer is posed on the Lie Group of $\mathbb{SLAM}_{n}\left(3\right)$
to mimic the true stochastic SLAM dynamics. The proposed approach
considers the velocity measurements to be attached with an unknown
bias and an unknown Gaussian noise. The proposed SLAM observer ensures
that the closed loop error signals are semi-globally uniformly ultimately
bounded. Simulation results demonstrates the efficiency and robustness
of the proposed approach, revealing its ability to localize the unknown
vehicle, as well as mapping the unknown environment given measurements
obtained from low-cost units.
\end{abstract}

%%%%%%%%%%%%%%%%%%%%%%%%%%%%%%%%%%%%%%%%%%%%%%%%%%%%%%%%%%%%%%%%%%%%%%%%%%%%%%%%
\section{Introduction}

Navigation is an essential part of robotics and control applications
\cite{hashim2021ACC,hashim2021_COMP_ENG_PRAC}. Successful navigation
of a vehicle in three dimensional (3D) space requires an accurate
estimation of its pose (\textit{i.e}., attitude and position) as well
as a map of the environment. The estimation of a vehicle's pose and
mapping of the environment is known as simultaneous localization and
mapping (SLAM). SLAM related applications are indispensable in indoor
and outdoor applications, especially in harsh environments. Over the
last twenty years, SLAM estimation has been studied extensively \cite{guo2020real,durrant2006simultaneous,Hashim2021AESCTE,sazdovski2015implicit,hashim2020LetterSLAM,zlotnik2018SLAM,li2018autonomous,milford2008mapping,sim2007study}.
SLAM estimation is accomplished using a group of sensor measurements,
where the sensors are attached to the body of the vehicle. The price
of a vehicle drops significantly in case of using low-cost sensing
units, but unfortunately, low-cost sensors are attached with high
levels of uncertainties, which compromise the estimation process.
Therefore, robust observers are necessary to compensate for the uncertainties
and to produce a reasonable estimate of the vehicle's pose, as well
as features of the environment.

In the past, the SLAM estimation problem have been addressed using
classical approaches that are commonly known as Gaussian filters \cite{hashim2021T_SMCS_SLAM}.
Examples include; the monoSLAM with object recognition using real-time
single camera \cite{castle2010combining}, neuro-adaptive FastSLAM
approach \cite{li2015neural}, incremental SLAM with constrained optimization
\cite{bai2018robust}, data fusion real-time RGB-D SLAM \cite{whelan2015real},
compressed unscented Kalman filter \cite{cheng2014compressed}, and
others. However, the SLAM problem is composed of two main parts: the
vehicle's pose and the features. The true feature dynamics are modeled
on the Lie Group of the Special Orthogonal Group $\mathbb{SO}\left(3\right)$,
while the vehicle's pose dynamics are modeled on the Lie Group of
the Special Euclidean Group $\mathbb{SE}\left(3\right)=\mathbb{SO}\left(3\right)\times\mathbb{R}^{3}$
\cite{hashim2021T_SMCS_SLAM,hashim2020TITS_SLAM}. Hence, the true
SLAM problem is highly nonlinear posed on the Lie Group $\mathbb{SLAM}_{n}\left(3\right)=\mathbb{SE}\left(3\right)\times\mathbb{R}^{3}\times\cdots\times\mathbb{R}^{3}$
which is not the unique source of complexity. Therefore, the SLAM
problem is better addressed on the Lie Group of $\mathbb{SLAM}_{n}\left(3\right)$
\cite{hashim2020LetterSLAM,zlotnik2018SLAM}. Over the last few years,
nonlinear filters for SLAM have increasingly become a perfect alternative
to supplant Gaussian filters. Examples of such include nonlinear filters
that rely on the measurements of angular velocity, translational velocity,
and features \cite{hashim2020LetterSLAM,zlotnik2018SLAM}. Other nonlinear
filters have also been proposed rely on the previously mentioned measurements,
as well as the inertial measurement unit (IMU) attached to the rigid-body
of the vehicle \cite{hashim2021T_SMCS_SLAM,hashim2020TITS_SLAM,nielsen2018ground,Hashim2021AESCTE}.
The solutions in \cite{hashim2020LetterSLAM,zlotnik2018SLAM,hashim2020TITS_SLAM,hashim2021T_SMCS_SLAM}
are nonlinear deterministic filters that compensate for unknown constant
bias attached to velocity measurements while the solution in \cite{Hashim2021AESCTE}
is a nonlinear stochastic filter compensates not only for the unknown
constant bias but also for random noise. It's worth noting that the
transient and steady-state error performance can be controlled using
the techniques in \cite{hashim2020LetterSLAM,hashim2020TITS_SLAM}.
To conclude, despite the fact that the SLAM has been addressed in
a stochastic sense, using stochastic differential equation in \cite{Hashim2021AESCTE},
the proposed algorithm relies on IMU data. This requirement increases
the computational cost.

In the present paper the SLAM problem is addressed on stochastic sense
on the Lie Group of $\mathbb{SLAM}_{n}\left(3\right)$, similar to
\cite{Hashim2021AESCTE}. Hence, the velocity data are assumed to
be corrupted with an unknown, constant bias and a Gaussian random
noise. Unlike \cite{Hashim2021AESCTE}, a nonlinear stochastic observer
for SLAM is proposed, capable of functioning without the need for
IMU data. The closed loop error signals are ensured to be semi-globally
uniformly ultimately bounded. 

After the above, the remainder of the paper is composed of four Sections.
Section \ref{sec:SE3_Problem-Formulation} presents the preliminaries
of $\mathbb{SO}\left(3\right)$ and $\mathbb{SE}\left(3\right)$,
the true SLAM dynamics and measurements, and error criteria. In Section
\ref{sec:SE3_Problem-Formulation} a nonlinear stochastic estimator
for SLAM is proposed , along with it's stability analysis. In Section
\ref{sec:SE3_Simulations}. the effectiveness of the proposed SLAM
observer schemes is demonstrated. Finally, Section \ref{sec:SE3_Conclusion}
presents the concluding results.

\section{Problem Formulation}\label{sec:SE3_Problem-Formulation}

\begin{tabular}{ll}
	\textbf{Notation} & \tabularnewline
	$\mathbb{R}$ & set of real numbers\tabularnewline
	$\mathbb{R}^{p\times q}$ & real space of dimension $p$-by-$q$\tabularnewline
	$\mathbf{I}_{n}\in\mathbb{R}^{n\times n}$ & identity matrix\tabularnewline
	$\left\Vert \cdot\right\Vert $ & Euclidean norm of a vector\tabularnewline
	$\mathbb{SO}\left(3\right)$ & Special Orthogonal Group\tabularnewline
	$\mathbb{SE}\left(3\right)$ & Special Euclidean Group\tabularnewline
\end{tabular}

\subsection{Preliminaries }

The attitude of a vehicle is defined by $R\in\mathbb{SO}\left(3\right)$
where $\mathbb{SO}\left(3\right)$ denotes the Special Orthogonal
Group
\[
\mathbb{SO}\left(3\right)=\left\{ \left.R\in\mathbb{R}^{3\times3}\right|RR^{\top}=\mathbf{I}_{3}\text{, }{\rm det}\left(R\right)=+1\right\} 
\]
$\left[\cdot\right]_{\times}$ denotes skew symmetric of a component
such that for $n\in\mathbb{R}^{3}$, one has:
\[
\left[n\right]_{\times}=\left[\begin{array}{ccc}
0 & -n_{3} & n_{2}\\
n_{3} & 0 & -n_{1}\\
-n_{2} & n_{1} & 0
\end{array}\right]\in\mathfrak{so}\left(3\right),\hspace{1em}n=\left[\begin{array}{c}
n_{1}\\
n_{2}\\
n_{3}
\end{array}\right]
\]
Pose of a vehicle can be represented by
\begin{equation}
\boldsymbol{T}=\left[\begin{array}{cc}
R & P\\
0_{1\times3} & 1
\end{array}\right]\in\mathbb{SE}\left(3\right)\subset\mathbb{R}^{4\times4}\label{eq:T_SLAM}
\end{equation}
where $R\in\mathbb{SO}\left(3\right)$ denotes the vehicle's attitude,
$P\in\mathbb{R}^{3}$ denotes vehicle's position, $\boldsymbol{T}$
denotes the homogeneous transformation matrix, and $\mathbb{SE}\left(3\right)$
is the Special Euclidean Group given by
\[
\mathbb{SE}\left(3\right)=\left\{ \left.\boldsymbol{T}=\left[\begin{array}{cc}
R & P\\
0_{1\times3} & 1
\end{array}\right]\right|R\in\mathbb{SO}\left(3\right),P\in\mathbb{R}^{3}\right\} 
\]

\subsection{Dynamics and Measurements}

\begin{figure*}
	\centering{}\includegraphics[scale=0.55]{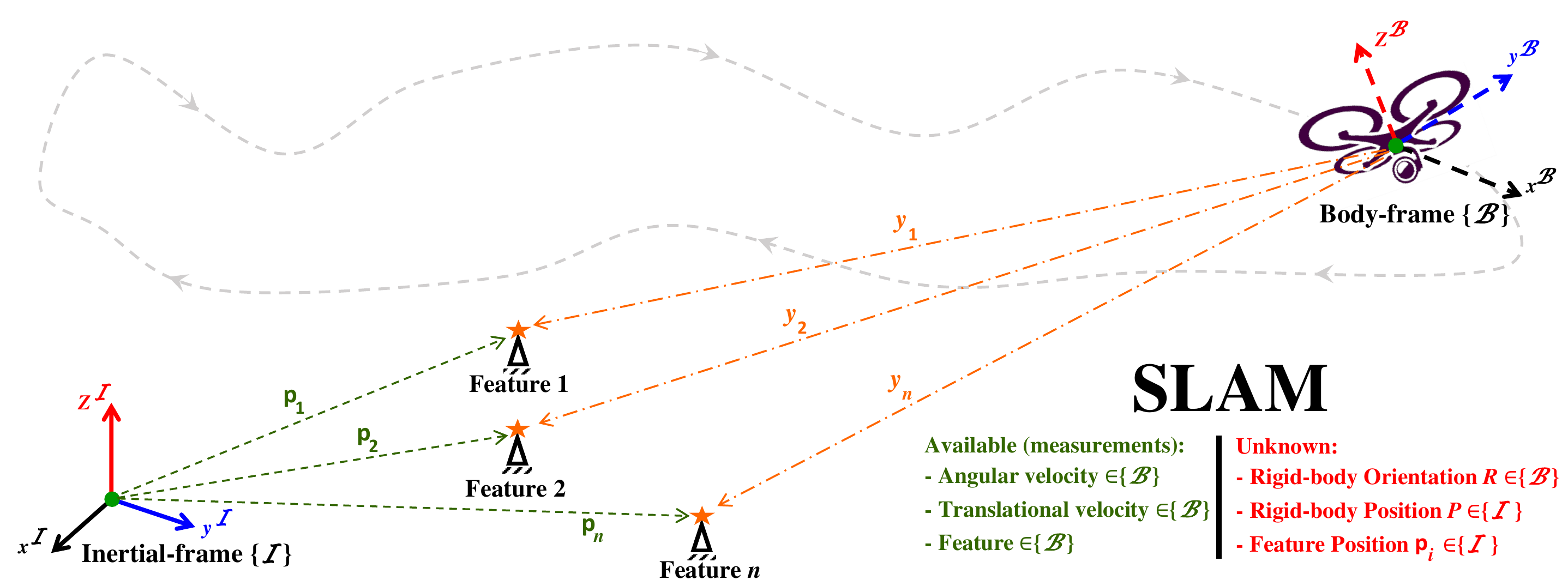}\caption{SLAM estimation problem \cite{hashim2020LetterSLAM}.}
	\label{fig:SLAM}
\end{figure*}

The SLAM problem considers a vehicle, whose pose is unknown, navigating
in an unknown environment. The unknown environment can be defined
through $n$ features. The vehicle's pose is denoted by $\boldsymbol{T}\in\mathbb{SE}\left(3\right)$
and ${\rm p}_{i}\in\mathbb{R}^{3}$. Fig. \ref{fig:SLAM} shows the
SLAM estimation problem in a 3D space.

The true dynamics of the vehicle's pose and the $i$th feature can
be described by \cite{hashim2020LetterSLAM}
\begin{align}
\left[\begin{array}{cc}
\dot{R} & \dot{P}\\
0_{1\times3} & 0
\end{array}\right] & =\left[\begin{array}{cc}
R & P\\
0_{1\times3} & 1
\end{array}\right]\left[\begin{array}{cc}
\left[\Omega\right]_{\times} & V\\
0_{1\times3} & 0
\end{array}\right]\label{eq:SLAM_T_dot}\\
\dot{{\rm p}}_{i} & =R{\rm v}_{i},\hspace{1em}\forall i=1,2,\ldots,n\label{eq:SLAM_p_dot}
\end{align}
or to put simply
\[
\begin{cases}
\dot{R} & =R\left[\Omega\right]_{\times}\\
\dot{P} & =RV\\
\dot{{\rm p}}_{i} & =R{\rm v}_{i},\hspace{1em}\forall i=1,2,\ldots,n
\end{cases}
\]
where $\Omega\in\mathbb{R}^{3}$ denotes the vehicle's angular velocity,
while $V\in\mathbb{R}^{3}$ denotes the vehicle's translational velocity.
It is worth noting that the SLAM dynamics in \eqref{eq:SLAM_T_dot}
and \eqref{eq:SLAM_p_dot} are posed on the Lie Group of $\mathbb{SLAM}_{n}\left(3\right)$,
see \cite{hashim2020LetterSLAM,hashim2020TITS_SLAM}. Let $n$ features
be available for measurement in the vehicle's frame (body-frame),
which can be obtained by a local vision unit. The $i$th measurement
is described by \cite{hashim2020SE3Stochastic}:
\begin{equation}
y_{i}=R^{\top}({\rm p}_{i}-P)+b_{i}^{y}+n_{i}^{y}\in\mathbb{R}^{3}\label{eq:SLAM_Vec_Landmark}
\end{equation}
for $i=1,2,\ldots,n$.

\begin{assum}\label{Assumption:Feature}At least 3 measured, non-collinear
	features ($n=3$) define a plane that's available at every time instant.\end{assum}

Angular velocity measurements are given by \cite{hashim2019SO3Wiley,Hashim2021AESCTE,odry2018kalman,odry2021open}:
\begin{equation}
\Omega_{m}=\Omega+b_{\Omega}+n_{\Omega}\in\mathbb{R}^{3}\label{eq:SLAM_Om_Meas}
\end{equation}
where $b_{\Omega}$ denotes a constant bias and $n_{\Omega}$ describes
an unknown noise. Likewise, translational velocity measurement is
described by \cite{hashim2020SE3Stochastic}:
\begin{equation}
V_{m}=V+b_{V}+n_{V}\in\mathbb{R}^{3}\label{eq:SLAM_V_Meas}
\end{equation}
where $b_{V}$ denotes a constant bias and $n_{V}$ describes an
unknown noise. Unknown uncertainties presenting challenge in variety
of applications \cite{Hashim2021AESCTE,hashim2019SO3Wiley,eltoukhy2018joint,eltoukhy2019robust}.

\subsection{Dynamics in Stochastic Sense}

\noindent Let $\left\{ n_{\Omega},t\geq t_{0}\right\} $ and $\left\{ n_{V},t\geq t_{0}\right\} $
be vector representations of independent, Brownian motion processes
\cite{ito1984lectures,hashim2019SO3Wiley,deng2001stabilization,hashim2020SE3Stochastic}:
\begin{equation}
n_{\Omega}=\mathcal{Q}_{\Omega}\frac{d\beta_{\Omega}}{dt},\hspace{1em}n_{V}=\mathcal{Q}_{V}\frac{d\beta_{V}}{dt}\label{eq:SLAM_noise}
\end{equation}
where $\mathcal{Q}_{\Omega}\in\mathbb{R}^{3\times3}$ and $\mathcal{Q}_{V}\in\mathbb{R}^{3\times3}$
refer to an unknown nonzero non-negative diagonal matrix. Note that
$\mathcal{Q}_{\Omega}$ and $\mathcal{Q}_{V}$ are bounded and time-variant.
$\mathcal{Q}_{\Omega}^{2}=\mathcal{Q}_{\Omega}\mathcal{Q}_{\Omega}^{\top}$
and $\mathcal{Q}_{V}^{2}=\mathcal{Q}_{V}\mathcal{Q}_{V}^{\top}$ denote
the covariance associated with the noises $n_{\Omega}$ and $n_{V}$,
respectively. It is worth noting that $\mathbb{P}\left\{ \beta_{\Omega}\left(0\right)=0\right\} =1,$
$\mathbb{P}\left\{ \beta_{V}\left(0\right)=0\right\} =1,$ $\mathbb{E}\left[d\beta_{\Omega}/dt\right]=\mathbb{E}\left[\beta_{\Omega}\right]=0$,
and $\mathbb{E}\left[d\beta_{V}/dt\right]=\mathbb{E}\left[\beta_{V}\right]=0$
such that $\mathbb{P}\left\{ \cdot\right\} $ denotes the probability
of an element while $\mathbb{E}[\cdot]$ refers to the expected value
of an element \cite{hashim2018SO3Stochastic}. Thus, the dynamics
in \eqref{eq:SLAM_T_dot} and \eqref{eq:SLAM_p_dot} can be reformulated
as:
\begin{align}
dR & =R[\Omega_{m}-b_{\Omega}]_{\times}dt-R\left[\mathcal{Q}_{\Omega}d\beta_{\Omega}\right]_{\times}\label{eq:SLAM_Rdot_Stoch}\\
dP & =R(V_{m}-b_{V})dt-R\mathcal{Q}_{V}d\beta_{V}\label{eq:SLAM_Posdot_Stoch}\\
d{\rm p}_{i} & =R{\rm v}_{i}dt,\hspace{1em}\forall i=1,2,\ldots,n\label{eq:SLAM_pdot_Stoch}
\end{align}
Now, let us define $\sigma$ as:
\begin{equation}
\sigma=\left[\begin{array}{c}
{\rm max}\left\{ \mathcal{Q}_{\Omega\left(1,1\right)}^{2},\mathcal{Q}_{V\left(1,1\right)}^{2}\right\} \\
{\rm max}\left\{ \mathcal{Q}_{\Omega\left(2,2\right)}^{2},\mathcal{Q}_{V\left(2,2\right)}^{2}\right\} \\
{\rm max}\left\{ \mathcal{Q}_{\Omega\left(3,3\right)}^{2},\mathcal{Q}_{V\left(3,3\right)}^{2}\right\} 
\end{array}\right]\label{eq:SLAM_s_covariance}
\end{equation}
where ${\rm max}\left\{ \cdot\right\} $ denotes the maximum value
of the associated component.

\section{Stochastic Observer Design}

The objective of this Section is to propose a stochastic observer
on the Lie Group of $\mathbb{SLAM}_{n}\left(3\right)$, capable of
localizing the unknown vehicle's pose and map the unknown environment.
First, define the estimate of attitude, position, and the $i$th feature
as $\hat{R}$, $\hat{P}$, and $\hat{{\rm p}}_{i}$, respectively.
Let the error in the vehicle's pose be given as:
\begin{align}
\hat{\boldsymbol{T}}\boldsymbol{T}^{-1} & =\left[\begin{array}{cc}
\hat{R} & \hat{P}\\
0_{1\times3} & 1
\end{array}\right]\left[\begin{array}{cc}
R^{\top} & -R^{\top}P\\
0_{1\times3} & 1
\end{array}\right]\nonumber \\
& =\left[\begin{array}{cc}
\tilde{R} & \tilde{P}\\
0_{1\times3} & 1
\end{array}\right]\label{eq:SLAM_Te}
\end{align}
where $\tilde{R}=\hat{R}R^{\top}$ and $\tilde{P}=\hat{P}-\tilde{R}P$.
Then, let the error in the feature be defined as
\begin{equation}
\tilde{{\rm p}}_{i}=\hat{{\rm p}}_{i}-\tilde{R}{\rm p}_{i}\label{eq:SLAM_pe}
\end{equation}
Let $\hat{b}_{\Omega}$ be the estimate of $b_{\Omega}$, $\hat{b}_{V}$
be the estimate of $b_{V}$, and $\hat{\sigma}$ be the estimate of
$\sigma$. Define the error in bias as:
\begin{align}
\tilde{b}_{\Omega} & =b_{\Omega}-\hat{b}_{\Omega}\nonumber \\
\tilde{b}_{V} & =b_{V}-\hat{b}_{V}\label{eq:SLAM_be}
\end{align}
Then define the error in the upper-bound covariance as
\begin{align}
\tilde{\sigma}_{\Omega} & =\sigma-\hat{\sigma}\label{eq:SLAM_se}
\end{align}
And define the $i$th error component as
\begin{equation}
e_{i}=\hat{{\rm p}}_{i}-\hat{R}y_{i}-\hat{P}=\tilde{{\rm p}}_{i}-\tilde{P}\label{eq:SLAM_e}
\end{equation}
The stochastic error dynamics, which will be defined in the subsequent
subsection are equivalent to:
\begin{equation}
de_{i}=f(e_{i},\tilde{b})dt+g(e_{i})\mathcal{Q}d\beta\label{eq:SLAM_e_dot}
\end{equation}
where $\mathcal{Q}=\left[\begin{array}{cc}
\mathcal{Q}_{\Omega} & 0_{3\times3}\\
0_{3\times3} & \mathcal{Q}_{V}
\end{array}\right]\in\mathbb{R}^{6\times6}$ and $\beta=[\beta_{\Omega}^{\top},\beta_{V}^{\top}]^{\top}\in\mathbb{R}^{6}$.
\begin{defn}
	\cite{hashim2018SO3Stochastic,hashim2019SO3Wiley}\label{def:SE3STCH_2}
	For the stochastic system in \eqref{eq:SLAM_e_dot}, Let $V(e_{i})$
	be a twice differentiable cost function. The differential operator
	is as follows:
	\[
	\mathcal{L}V\left(e_{i}\right)=V_{e_{i}}^{\top}f+\frac{1}{2}{\rm Tr}\left\{ g\mathcal{Q}^{2}g^{\top}V_{e_{i}e_{i}}\right\} 
	\]
	with $V_{e_{i}}=\partial V/\partial e_{i}$ and $V_{e_{i}e_{i}}=\partial^{2}V/\partial e_{i}^{2}$.
\end{defn}
\begin{lem}
	\label{lem:SE3STCH_1} \cite{deng2001stabilization} For the stochastic
	dynamics in \eqref{eq:SLAM_e_dot} let $V(e_{i})$ be a twice differentiable
	cost function. Define $\bar{\upsilon}_{1}(\cdot)$ and $\bar{\upsilon}_{2}(\cdot)$
	as class $\mathcal{K}_{\infty}$ functions and define $\boldsymbol{{\rm c}}>0$
	and $\mathbf{k}\geq0$ as scalars. Let $\boldsymbol{\mathcal{N}}(||e_{i}||)$
	be a non-negative function where,
	\begin{equation}
	\bar{\upsilon}_{1}(||e_{i}||)\leq V\leq\bar{\upsilon}_{2}(||e_{i}||)\label{eq:SE3STCH_Vfunction_Lyap}
	\end{equation}
	\begin{align}
	\mathcal{L}V(e_{i})= & V_{e_{i}}^{\top}f+\frac{1}{2}{\rm Tr}\left\{ g\mathcal{Q}^{2}g^{\top}V_{e_{i}e_{i}}\right\} \nonumber \\
	\leq & -\boldsymbol{{\rm c}}\boldsymbol{\mathcal{N}}(||e_{i}||)+\mathbf{k}\label{eq:SE3STCH_dVfunction_Lyap}
	\end{align}
	Then for $e_{i}(0)\in\mathbb{R}^{3}$, there is an almost unique strong
	solution on $\left[0,\infty\right)$ for \eqref{eq:SLAM_e_dot}. Also,
	$e_{i}$ is bounded in probability, following the inequality below:
	\begin{equation}
	\mathbb{E}\left[V\left(e_{i}\right)\right]\leq V\left(e_{i}\left(0\right)\right){\rm exp}\left(-\boldsymbol{{\rm c}}t\right)+\frac{\mathbf{k}}{\boldsymbol{{\rm c}}}\label{eq:SE3STCH_EVfunction_Lyap}
	\end{equation}
	and $e_{i}$ is semi-globally uniformly ultimately bounded.
\end{lem}
Consider the following nonlinear stochastic observer on the Lie Group
of $\mathbb{SLAM}_{n}\left(3\right)$:

\begin{align}
\dot{\hat{\boldsymbol{T}}}= & \hat{\boldsymbol{T}}\left[\begin{array}{cc}
\left[\Omega_{m}-\hat{b}_{\Omega}-W_{\Omega}\right]_{\times} & V_{m}-\hat{b}_{V}-W_{V}\\
0_{1\times3} & 0
\end{array}\right]\label{eq:SLAM_T_est_dot_f2}\\
\dot{{\rm \hat{p}}}_{i}= & -\left(k_{p}+\frac{5}{\alpha_{i}}\hat{\sigma}+\frac{3}{\varrho\alpha_{i}}(1-{\rm Tr}\{[\hat{{\rm p}}_{i}]_{\times}^{2}\})^{2}\right)e_{i}\label{eq:SLAM_p_est_dot_f2}\\
\left[\begin{array}{c}
W_{\Omega}\\
W_{V}
\end{array}\right]= & \sum_{i=1}^{n}k_{w}\left[\begin{array}{c}
-\hat{R}^{\top}\left[\hat{R}y_{i}+\hat{{\rm p}}_{i}\right]_{\times}e_{i}\\
\hat{R}^{\top}\left(\left[\hat{P}\right]_{\times}\left[\hat{R}y_{i}+\hat{{\rm p}}_{i}\right]_{\times}-\mathbf{I}_{3}\right)e_{i}
\end{array}\right]\label{eq:SLAM_W_f2}\\
\left[\begin{array}{c}
\dot{\hat{b}}_{\Omega}\\
\dot{\hat{b}}_{V}
\end{array}\right]= & -\sum_{i=1}^{n}\frac{\Gamma}{\alpha_{i}}\left[\begin{array}{c}
\hat{R}^{\top}\left[\hat{R}y_{i}+\hat{{\rm p}}_{i}-\hat{P}\right]_{\times}e_{i}\\
\hat{R}^{\top}e_{i}
\end{array}\right]\nonumber \\
& -k_{b}\Gamma\left[\begin{array}{c}
\hat{b}_{\Omega}\\
\hat{b}_{V}
\end{array}\right]\label{eq:SLAM_b_est_dot_f2}\\
\dot{\hat{\sigma}}= & 5\sum_{i=1}^{n}\frac{\gamma_{\sigma}}{\alpha_{i}^{2}}\left\Vert e_{i}\right\Vert ^{4}-k_{\sigma}\gamma_{\sigma}\hat{\sigma}\label{eq:SLAM_s_est_dot_f2}
\end{align}
where $W_{\Omega},W_{V}\in\mathbb{R}^{3}$ are correction factors,
$\hat{b}_{\Omega},\hat{b}_{V}\in\mathbb{R}^{3}$ denote bias estimates,
and $\hat{\sigma}\in\mathbb{R}^{3}$ denotes the upper-bound covariance
estimate. Also, $k_{\sigma}$, $\gamma_{\sigma}$, $k_{b}$, $k_{w}$,
$\Gamma$, $\varrho$, and $\alpha_{i}$ are positive constants.
\begin{thm}
	\label{thm:SE3STCH_1}\textbf{ }Consider the stochastic dynamics in
	\eqref{eq:SLAM_Rdot_Stoch}-\eqref{eq:SLAM_pdot_Stoch}. Let the stochastic
	observer in \eqref{eq:SLAM_T_est_dot_f2}-\eqref{eq:SLAM_s_est_dot_f2}
	be coupled with the velocity measurements in \eqref{eq:SLAM_Om_Meas}
	and \eqref{eq:SLAM_V_Meas} and the error vectors in \eqref{eq:SLAM_e}.
	Then, suppose that Assumption \ref{Assumption:Feature} holds true
	and the design parameters $k_{\sigma}$, $\gamma_{\sigma}$, $k_{b}$,
	$k_{w}$, $\Gamma$, $\varrho$, and $\alpha_{i}$ are selected as
	positive constants. All the closed-loop error signals are semi-globally
	uniformly ultimately bounded.
\end{thm}
\textbf{Proof. }From the true stochastic dynamics in \eqref{eq:SLAM_Rdot_Stoch}-\eqref{eq:SLAM_pdot_Stoch},
the stochastic observer design in \eqref{eq:SLAM_T_est_dot_f2}-\eqref{eq:SLAM_s_est_dot_f2},
and the error definitions in \eqref{eq:SLAM_Te}-\eqref{eq:SLAM_e},
one finds that:{\footnotesize{}
	\begin{align*}
	de_{i}= & -\left[\begin{array}{c}
	\left[\hat{R}y_{i}+\hat{P}\right]_{\times}\\
	\mathbf{I}_{3}
	\end{array}\right]^{\top}\left[\begin{array}{cc}
	\hat{R} & 0_{3\times3}\\
	\left[\hat{P}\right]_{\times}\hat{R} & \hat{R}
	\end{array}\right]\left[\begin{array}{c}
	\tilde{b}_{\Omega}-W_{\Omega}\\
	\tilde{b}_{\Omega}-W_{V}
	\end{array}\right]dt\\
	& +d\hat{{\rm p}}_{i}-\left[\begin{array}{c}
	\left[\hat{R}y_{i}+\hat{P}\right]_{\times}\\
	\mathbf{I}_{3}
	\end{array}\right]^{\top}\left[\begin{array}{cc}
	\hat{R} & 0_{3\times3}\\
	\left[\hat{P}\right]_{\times}\hat{R} & \hat{R}
	\end{array}\right]\mathcal{Q}d\beta
	\end{align*}
}such that
\begin{align}
de_{i}= & f(e_{i},\tilde{b}_{\Omega},\tilde{b}_{V})dt+g(e_{i})\mathcal{Q}d\beta\label{eq:SE3STCH_Rt_dot}
\end{align}
For $V:=V(e,\tilde{b}_{\Omega},\tilde{b}_{V},\tilde{\sigma})$, define
the following Lyapunov candidate function:
\begin{align}
V= & \sum_{i=1}^{n}\frac{1}{4\alpha_{i}}\left\Vert e_{i}\right\Vert ^{4}+\frac{1}{2}\tilde{b}_{\Omega}^{\top}\Gamma^{-1}\tilde{b}_{\Omega}+\frac{1}{2}\tilde{b}_{V}^{\top}\Gamma^{-1}\tilde{b}_{V}\nonumber \\
& +\frac{1}{2\gamma_{\sigma}}\left\Vert \tilde{\sigma}\right\Vert ^{2}\label{eq:SE3STCH_V}
\end{align}
where $e_{i}$ is defined in \eqref{eq:SLAM_e}, $\tilde{b}_{\Omega}$
and $\tilde{b}_{V}$ are defined in \eqref{eq:SLAM_be}, and $\tilde{\sigma}$
is defined in \eqref{eq:SLAM_se}. Based on Definition \ref{def:SE3STCH_2},
the differential operator $\mathcal{L}V$ is equivalent to:
\begin{align}
\mathcal{L}V= & V_{e}^{\top}f+\frac{1}{2}{\rm Tr}\left\{ g\mathcal{Q}^{2}g^{\top}V_{ee}\right\} -\tilde{b}_{\Omega}^{\top}\Gamma^{-1}\dot{\hat{b}}_{\Omega}\nonumber \\
& -\tilde{b}_{V}^{\top}\Gamma^{-1}\dot{\hat{b}}_{V}-\frac{1}{\gamma_{\sigma}}\tilde{\sigma}\dot{\hat{\sigma}}\label{eq:SE3STCH_VL1}
\end{align}
such that
\begin{align}
\mathcal{L}V\leq & \sum_{i=1}^{n}\frac{1}{\alpha_{i}}\left\Vert e_{i}\right\Vert ^{2}e_{i}^{\top}\frac{d}{dt}\hat{{\rm p}}_{i}\nonumber \\
& -\sum_{i=1}^{n}\frac{1}{\alpha_{i}}\left\Vert e_{i}\right\Vert ^{2}e_{i}^{\top}\left[\begin{array}{c}
\left[\hat{R}y_{i}+\hat{P}\right]_{\times}\\
\mathbf{I}_{3}
\end{array}\right]^{\top}\nonumber \\
& \hspace{6em}\times\left[\begin{array}{cc}
\hat{R} & 0_{3\times3}\\
\left[\hat{P}\right]_{\times}\hat{R} & \hat{R}
\end{array}\right]\left[\begin{array}{c}
\tilde{b}_{\Omega}-W_{\Omega}\\
\tilde{b}_{\Omega}-W_{V}
\end{array}\right]\nonumber \\
& +\sigma\sum_{i=1}^{n}\frac{9}{2\alpha_{i}^{2}}\left(\left\Vert e_{i}\right\Vert ^{4}+(1-{\rm Tr}\{[\hat{{\rm p}}_{i}]_{\times}^{2}\})\left\Vert e_{i}\right\Vert ^{2}\right)\nonumber \\
& -\tilde{b}_{\Omega}^{\top}\Gamma^{-1}\dot{\hat{b}}_{\Omega}-\tilde{b}_{V}^{\top}\Gamma^{-1}\dot{\hat{b}}_{V}-\frac{1}{\gamma_{\sigma}}\tilde{\sigma}\dot{\hat{\sigma}}\label{eq:SE3STCH_VL2}
\end{align}
\begin{align}
& \mathcal{L}V\leq-\sum_{i=1}^{n}\frac{1}{\alpha_{i}}\left\Vert e_{i}\right\Vert ^{2}e_{i}^{\top}\left(\mathbf{I}_{3}-\left[\hat{R}y_{i}+\hat{P}\right]_{\times}^{2}\right)e_{i}\nonumber \\
& -\sum_{i=1}^{n}\left(\frac{k_{p}}{\alpha_{i}}+\frac{\sigma}{2\alpha_{i}^{2}}+\frac{3}{4\alpha_{i}^{2}\varrho}({\rm Tr}\{[\hat{{\rm p}}_{i}]_{\times}^{2}\}+1)^{2}\right)\left\Vert e_{i}\right\Vert ^{4}\nonumber \\
& +\frac{9\varrho\sigma^{2}}{4\sum_{i=1}^{n}\alpha_{i}^{2}}+k_{b}\tilde{b}_{\Omega}^{\top}\left(b_{\Omega}-\tilde{b}_{\Omega}\right)\nonumber \\
& +k_{b}\tilde{b}_{V}^{\top}\left(b_{V}-\tilde{b}_{V}\right)+k_{\sigma}\tilde{\sigma}\left(\sigma-\tilde{\sigma}\right)\label{eq:SE3STCH_VL3}
\end{align}
In view of Young's inequality one has
\begin{align*}
k_{b}\tilde{b}_{\Omega}^{\top}b_{\Omega} & \leq\frac{k_{b}}{2}\left\Vert b_{\Omega}\right\Vert ^{2}+\frac{k_{b}}{2}\left\Vert \tilde{b}_{\Omega}\right\Vert ^{2}\\
k_{b}\tilde{b}_{V}^{\top}b_{V} & \leq\frac{k_{b}}{2}\left\Vert b_{V}\right\Vert ^{2}+\frac{k_{b}}{2}\left\Vert \tilde{b}_{V}\right\Vert ^{2}\\
k_{\sigma}\tilde{\sigma}^{\top}\sigma & \leq\frac{k_{\sigma}}{2}\left\Vert \sigma\right\Vert ^{2}+\frac{k_{\sigma}}{2}\left\Vert \tilde{\sigma}\right\Vert ^{2}
\end{align*}
Now, let's define the following variables:
\[
K_{i}=4k_{p}+\frac{2\sigma}{\alpha_{i}}+\frac{3}{\alpha_{i}\varrho},\hspace{1em}\forall i=1,2,\ldots,n
\]
\[
\mathcal{H}=\left[\begin{array}{c|c}
\begin{array}{ccc}
K_{1}\mathbf{I}_{3} & \cdots & 0_{3\times3}\\
\vdots & \ddots & \vdots\\
0_{3\times3} & \cdots & K_{n}\mathbf{I}_{3}
\end{array} & 0_{3n\times7}\\
\hline 0_{7\times3n} & \begin{array}{cc}
\frac{1}{2}k_{b}\Gamma & 0_{6\times1}\\
0_{1\times6} & \frac{1}{2}k_{\sigma}\gamma_{\sigma}
\end{array}
\end{array}\right]
\]
\[
\tilde{Y}=\left[\frac{e_{1}^{\top}}{2\sqrt{\alpha_{1}}},\ldots,\frac{e_{n}^{\top}}{2\sqrt{\alpha_{n}}},\tilde{b}_{U}^{\top}\sqrt{\frac{\Gamma^{-1}}{2}},\frac{\tilde{\sigma}}{\sqrt{2\gamma_{\sigma}}}\right]^{\top}
\]
\[
\eta_{2}=\frac{k_{b}}{2}\left\Vert b_{U}\right\Vert ^{2}+\left(\frac{9\varrho}{4\sum_{i=1}^{n}\alpha_{i}^{2}}+\frac{k_{\sigma}}{2}\right)\left\Vert \sigma\right\Vert ^{2}
\]
where $\mathcal{H}\in\mathbb{R}^{(3n+7)\times(3n+7)}$ and $\tilde{Y}\in\in\mathbb{R}^{(3n+7)\times1}$.
Thereby, $\mathcal{L}V$ in \eqref{eq:SE3STCH_VL3} can be rewritten
as:
\begin{align}
\mathcal{L}V\leq & -f(||e_{i}||^{2})-\tilde{Y}^{\top}\mathcal{H}\tilde{Y}+\eta_{2}\label{eq:SE3STCH_VL_Final-1}
\end{align}
such that:
\begin{equation}
\mathcal{L}V\leq-\lambda_{\min}(\mathcal{H})V+\eta_{2}\label{eq:SE3STCH_VL_Final-1-1}
\end{equation}
where $\lambda_{\min}=\lambda_{\min}(\mathcal{H})$ denotes the minimum
eigenvalue of $\mathcal{H}$. Hence, it can be shown that:
\begin{align}
\frac{d\left(\mathbb{E}\left[V\right]\right)}{dt}=\mathbb{E}\left[\mathcal{L}V\right] & \leq-\lambda_{\min}\mathbb{E}\left[V\right]+\eta_{2}\label{eq:SE3STCH_VL_Final-2}
\end{align}
and utilizing Lemma \ref{lem:SE3STCH_1}, one obtains the following
inequality:
\begin{align}
0\leq\mathbb{E}\left[V\left(t\right)\right] & \leq V\left(0\right)\exp\left(-\lambda_{\min}t\right)+\frac{\eta_{2}}{\lambda_{\min}},\forall t\geq0\label{eq:SE3STCH_V_Final-1}
\end{align}
Consequently, $\tilde{Y}$ is semi-globally uniformly ultimately bounded
completing the proof.

\section{Simulation}\label{sec:SE3_Simulations}

This Section shows the robustness of the proposed stochastic observer
for SLAM. The observer is tested against high levels of uncertainties,
corrupting the velocity and feature measurements. Let the true attitude
and position of the vehicle be defined as:
\[
R\left(0\right)=\mathbf{I}_{3},\hspace{1em}P\left(0\right)=[0,0,1]^{\top}
\]
and consider the true angular and translational velocities to be
$\Omega=[0,0,0.1]^{\top}({\rm rad/sec})$ and $V=[1.5,0,0]^{\top}({\rm m/sec})$,
respectively. Let four, non-collinear features be distributed in the
map relative to the inertial-frame, where ${\rm p}_{1}=[1.5,0,0]^{\top}$,
${\rm p}_{2}=[-1.5,0,0]^{\top}$, ${\rm p}_{3}=[0,1.5,0]^{\top}$,
and ${\rm p}_{4}=[0,-1.5,0]^{\top}$. Consider the measurements of
angular velocities to be corrupted with an unknown, constant bias
and a random noise where $b_{\Omega}=[0.05,-0.06,-0.07]^{\top}({\rm rad/sec})$
and $b_{V}=[0.04,0.06,-0.08]^{\top}({\rm m/sec})$, $n_{\Omega}=\mathcal{N}\left(0,0.1\right)$ $({\rm rad/sec})$,
and $n_{V}=\mathcal{N}\left(0,0.12\right)({\rm m/sec})$. Let the
initial estimate of the vehicle's pose be:
\[
\hat{R}\left(0\right)=\mathbf{I}_{3},\hspace{1em}\hat{P}\left(0\right)=[0,0,0]^{\top}
\]
and consider the four feature estimates to be, $\hat{{\rm p}}_{1}\left(0\right)=\hat{{\rm p}}_{2}\left(0\right)=\hat{{\rm p}}_{3}\left(0\right)=\hat{{\rm p}}_{4}\left(0\right)=[0,0,0]^{\top}$.
Let the design parameters be chosen as $k_{\sigma}=1$, $\gamma_{\sigma}=1$,
$k_{b}=10$, $k_{w}=10$, $\Gamma=5\mathbf{I}_{3}$, $\varrho=0.3$,
and $\alpha_{i}=0.04$ for $i=1,2,3,4$. Also, let chose the initial
estimates to be, $\hat{b}_{\Omega}\left(0\right)=\hat{b}_{V}\left(0\right)=[0,0,0]^{\top}$
and $\hat{\sigma}\left(0\right)=[0,0,0]^{\top}$.

\begin{figure}[h]
	\centering{}\includegraphics[scale=0.27]{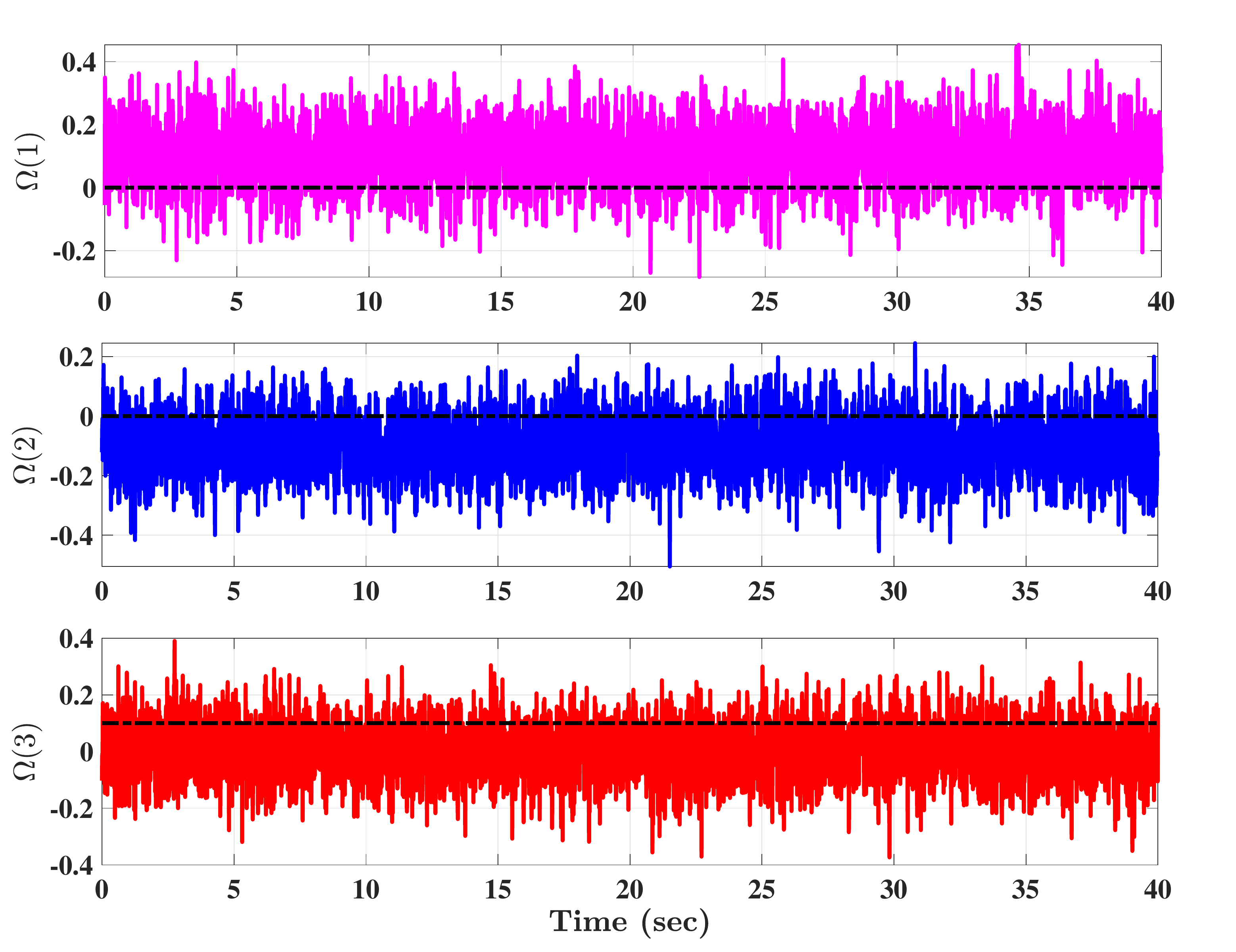}\caption{Angular velocity: true (black center-line) and measured (colored)}
	\label{fig:Fig4_VEL} 
\end{figure}

\begin{figure}[h]
	\centering{}\includegraphics[scale=0.27]{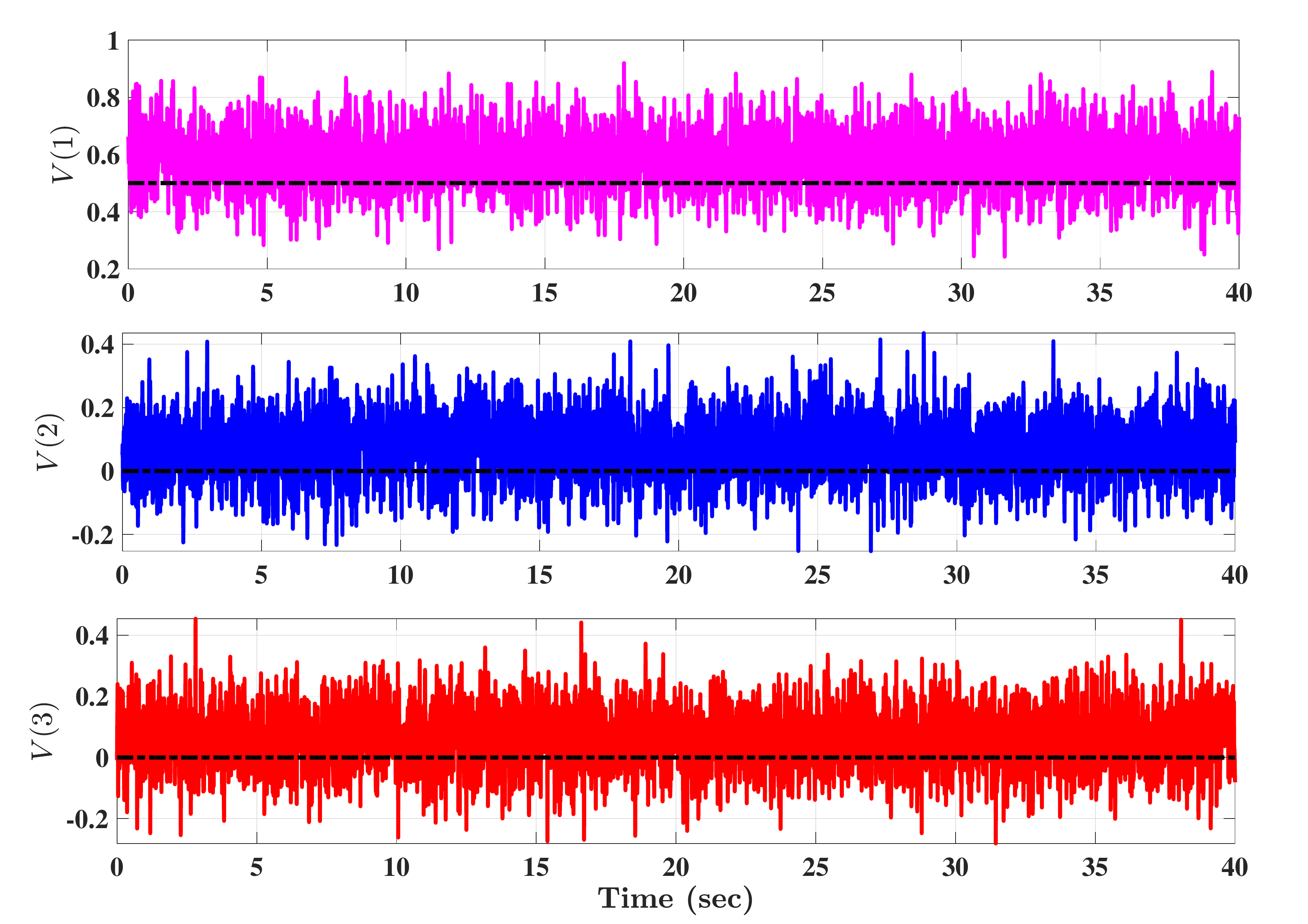}\caption{Translational velocity: true (black center-line) and measured (colored)}
	\label{fig:Fig4_VEL-1}
\end{figure}

\begin{figure}[h]
	\centering{}\includegraphics[scale=0.3]{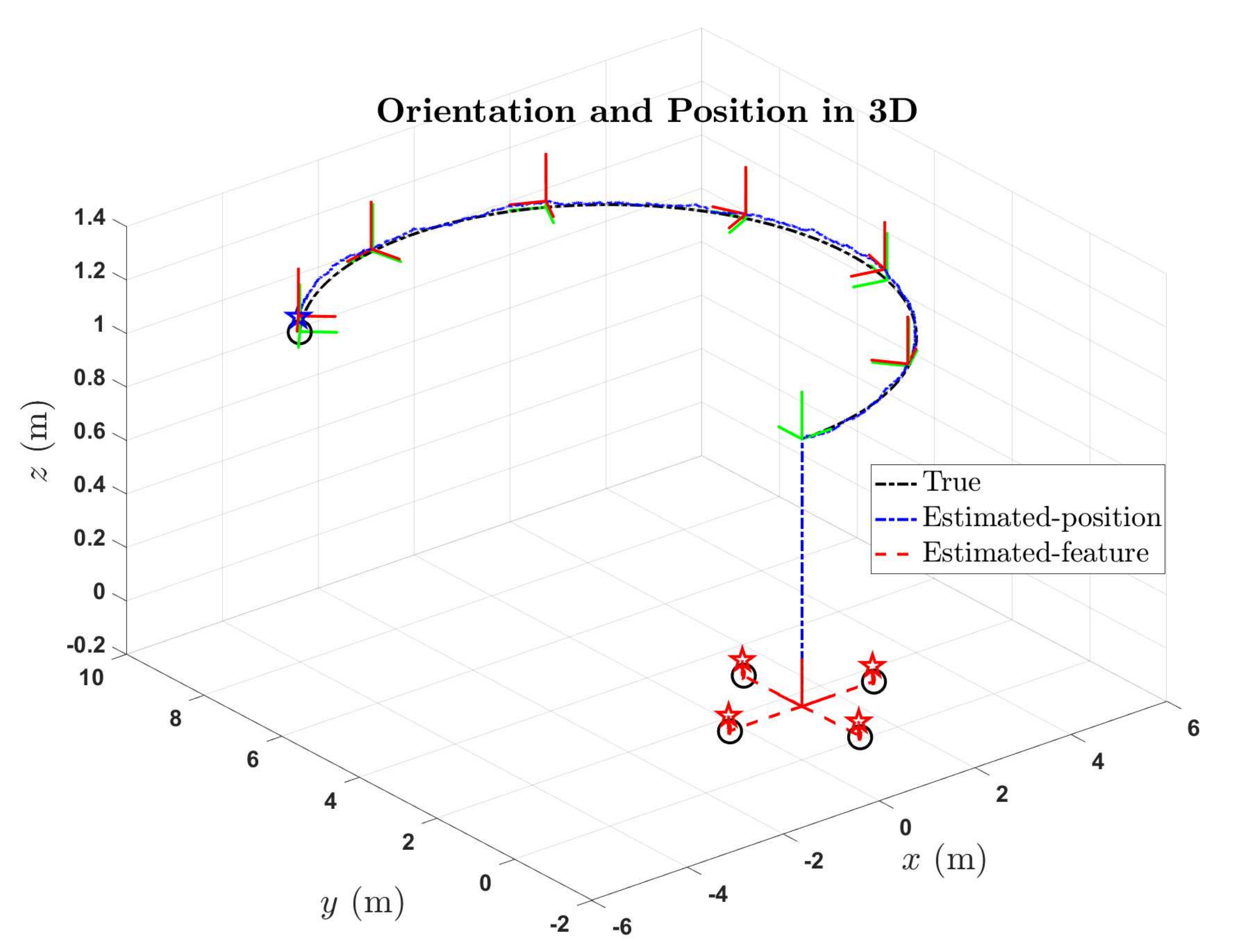}\caption{Simultaneous localization and mapping}
	\label{fig:Fig5_3D} 
\end{figure}

\begin{figure}[h]
	\centering{}\includegraphics[scale=0.32]{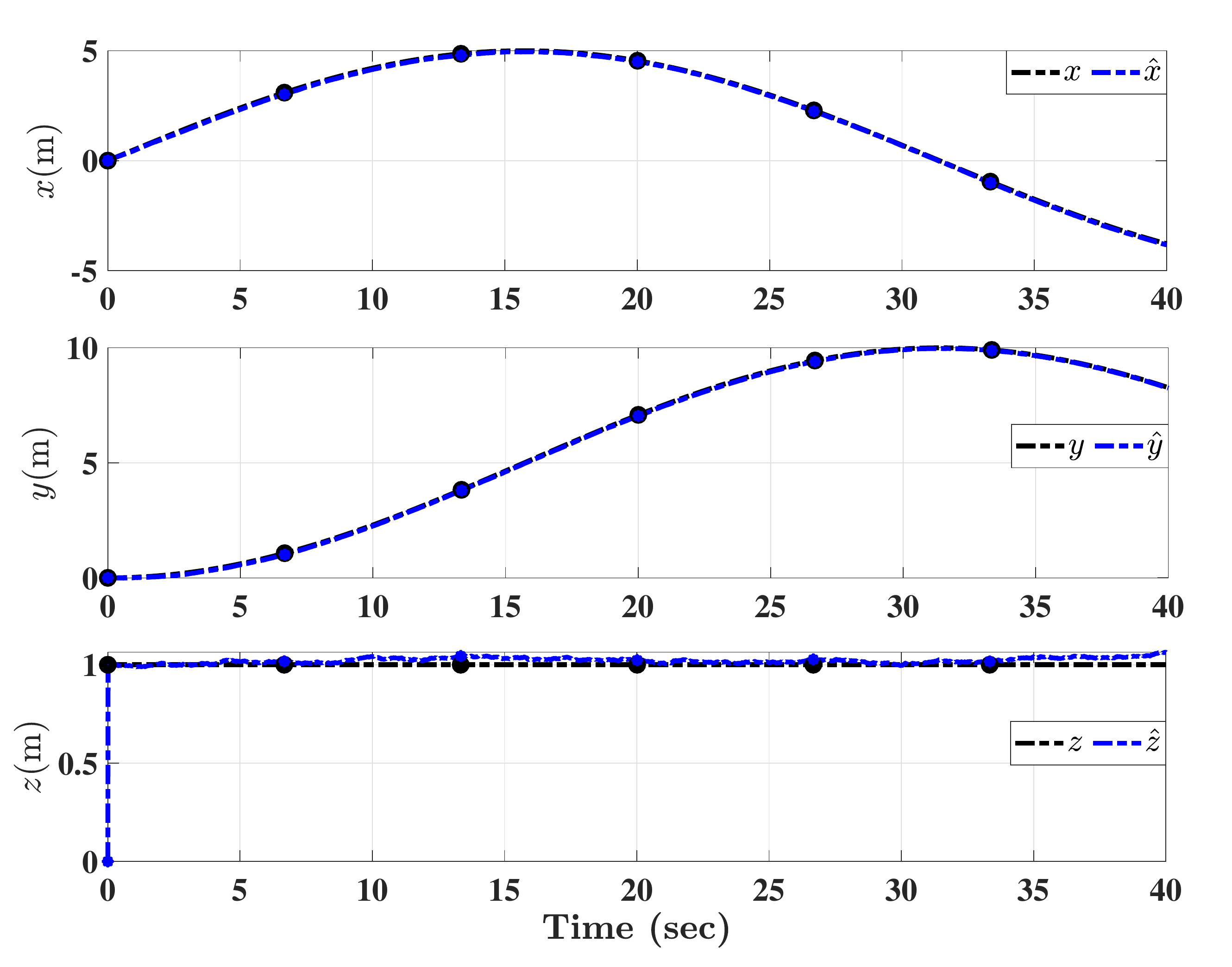}\caption{Position trajectory: x, y, and z}
	\label{fig:Fig6_P} 
\end{figure}

Fig. \ref{fig:Fig4_VEL} illustrates the high level of uncertainties
attached to the angular velocity and Fig. \ref{fig:Fig4_VEL-1} shows
the high levels of uncertainties attached to the translational velocity.
The uncertainties the unknown, constant bias and random noise. Fig.
\ref{fig:Fig5_3D} reveals strong and successful tracking performance
of the proposed stochastic observer to follow the true trajectory
starting from large error in initialization. As well, pose estimate
initiated at the origin, was set to the true pose trajectory in a
short period of time. Likewise, the feature estimates started at origin
and converted to the true features. Fig. \ref{fig:Fig6_P} shows the
strong tracking performance of the position estimate relative to the
true position.

\section{Conclusion}\label{sec:SE3_Conclusion}

In this paper, the SLAM estimation problem has been addressed in a
stochastic sense. A nonlinear stochastic observer posed on the Lie
Group of $\mathbb{SLAM}_{n}\left(3\right)$ has been proposed. The
proposed observer can operate using angular and translational velocity
measurements, along with the available feature measurements obtained
from a vision unit. It has been assumed that the angular and translational
velocity measurements are corrupted, not only with an unknown, constant
bias, but also with a random Gaussian noise. The closed loop error
signals have been shown to be semi-globally uniformly ultimately bounded.
Finally, simulation results showed the effectiveness and robustness
of the proposed approach given high level of uncertainties in the
measurements.

\section*{Acknowledgment}

The authors would like to thank \textbf{Maria Shaposhnikova} for proofreading
the article.

\bibliographystyle{IEEEtran}
\bibliography{IEEE_Conf_SLAM}

\begin{thebibliography}{10}
\providecommand{\url}[1]{#1}
\csname url@samestyle\endcsname
\providecommand{\newblock}{\relax}
\providecommand{\bibinfo}[2]{#2}
\providecommand{\BIBentrySTDinterwordspacing}{\spaceskip=0pt\relax}
\providecommand{\BIBentryALTinterwordstretchfactor}{4}
\providecommand{\BIBentryALTinterwordspacing}{\spaceskip=\fontdimen2\font plus
\BIBentryALTinterwordstretchfactor\fontdimen3\font minus
  \fontdimen4\font\relax}
\providecommand{\BIBforeignlanguage}[2]{{%
\expandafter\ifx\csname l@#1\endcsname\relax
\typeout{** WARNING: IEEEtran.bst: No hyphenation pattern has been}%
\typeout{** loaded for the language `#1'. Using the pattern for}%
\typeout{** the default language instead.}%
\else
\language=\csname l@#1\endcsname
\fi
#2}}
\providecommand{\BIBdecl}{\relax}
\BIBdecl

\bibitem{hashim2021ACC}
H.~A. {Hashim}, ``{GPS}-denied navigation: Attitude, position, linear velocity,
  and gravity estimation with nonlinear stochastic observer,'' in \emph{2021
  American Control Conference (ACC)}.\hskip 1em plus 0.5em minus 0.4em\relax
  IEEE, 2021, pp. 1146--1151.

\bibitem{hashim2021_COMP_ENG_PRAC}
H.~A. Hashim, M.~Abouheaf, and M.~A. Abido, ``Geometric stochastic filter with
  guaranteed performance for autonomous navigation based on {IMU} and feature
  sensor fusion,'' \emph{Control Engineering Practice}, vol.~PP, no.~PP, pp.
  1--11, 2021.

\bibitem{guo2020real}
J.~Guo, Y.~He, X.~Qi, G.~Wu, Y.~Hu, B.~Li, and J.~Zhang, ``Real-time
  measurement and estimation of the 3d geometry and motion parameters for
  spatially unknown moving targets,'' \emph{Aerospace Science and Technology},
  vol.~97, p. 105619, 2020.

\bibitem{durrant2006simultaneous}
H.~Durrant-Whyte and T.~Bailey, ``Simultaneous localization and mapping: part
  i,'' \emph{IEEE robotics \& automation magazine}, vol.~13, no.~2, pp.
  99--110, 2006.

\bibitem{Hashim2021AESCTE}
H.~A. Hashim, ``A geometric nonlinear stochastic filter for simultaneous
  localization and mapping,'' \emph{Aerospace Science and Technology}, vol.
  111, p. 106569, 2021.

\bibitem{sazdovski2015implicit}
V.~Sazdovski, A.~Kitanov, and I.~Petrovic, ``Implicit observation model for
  vision aided inertial navigation of aerial vehicles using single camera
  vector observations,'' \emph{Aerospace science and technology}, vol.~40, pp.
  33--46, 2015.

\bibitem{hashim2020LetterSLAM}
H.~A. Hashim, ``Guaranteed performance nonlinear observer for simultaneous
  localization and mapping,'' \emph{IEEE Control Systems Letters}, vol.~5,
  no.~1, pp. 91--96, 2021.

\bibitem{zlotnik2018SLAM}
D.~E. Zlotnik and J.~R. Forbes, ``Gradient-based observer for simultaneous
  localization and mapping,'' \emph{IEEE Transactions on Automatic Control},
  vol.~63, no.~12, pp. 4338--4344, 2018.

\bibitem{li2018autonomous}
M.~Li and B.~Xu, ``Autonomous orbit and attitude determination for earth
  satellites using images of regular-shaped ground objects,'' \emph{Aerospace
  Science and Technology}, vol.~80, pp. 192--202, 2018.

\bibitem{milford2008mapping}
M.~J. Milford and G.~F. Wyeth, ``Mapping a suburb with a single camera using a
  biologically inspired slam system,'' \emph{IEEE Transactions on Robotics},
  vol.~24, no.~5, pp. 1038--1053, 2008.

\bibitem{sim2007study}
R.~Sim, P.~Elinas, and J.~J. Little, ``A study of the rao-blackwellised
  particle filter for efficient and accurate vision-based slam,''
  \emph{International Journal of Computer Vision}, vol.~74, no.~3, pp.
  303--318, 2007.

\bibitem{hashim2021T_SMCS_SLAM}
H.~A. Hashim and A.~E.~E. Eltoukhy, ``Nonlinear filter for simultaneous
  localization and mapping on a matrix lie group using {IMU} and feature
  measurements,'' \emph{IEEE Transactions on Systems, Man, and Cybernetics:
  Systems}, vol.~PP, no.~PP, pp. 1--12, 2021.

\bibitem{castle2010combining}
R.~O. Castle, G.~Klein, and D.~W. Murray, ``Combining monoslam with object
  recognition for scene augmentation using a wearable camera,'' \emph{Image and
  Vision Computing}, vol.~28, no.~11, pp. 1548--1556, 2010.

\bibitem{li2015neural}
Q.-L. Li, Y.~Song, and Z.-G. Hou, ``Neural network based fastslam for
  autonomous robots in unknown environments,'' \emph{Neurocomputing}, vol. 165,
  pp. 99--110, 2015.

\bibitem{bai2018robust}
F.~Bai, T.~Vidal-Calleja, and S.~Huang, ``Robust incremental slam under
  constrained optimization formulation,'' \emph{IEEE Robotics and Automation
  Letters}, vol.~3, no.~2, pp. 1207--1214, 2018.

\bibitem{whelan2015real}
T.~Whelan, M.~Kaess, H.~Johannsson, M.~Fallon, J.~J. Leonard, and J.~McDonald,
  ``Real-time large-scale dense rgb-d slam with volumetric fusion,'' \emph{The
  International Journal of Robotics Research}, vol.~34, no. 4-5, pp. 598--626,
  2015.

\bibitem{cheng2014compressed}
J.~Cheng, J.~Kim, Z.~Jiang, and X.~Yang, ``Compressed unscented kalman
  filter-based slam,'' in \emph{2014 IEEE International Conference on Robotics
  and Biomimetics (ROBIO 2014)}.\hskip 1em plus 0.5em minus 0.4em\relax IEEE,
  2014, pp. 1602--1607.

\bibitem{hashim2020TITS_SLAM}
H.~A. Hashim and A.~E.~E. Eltoukhy, ``Landmark and {IMU} data fusion:
  Systematic convergence geometric nonlinear observer for {SLAM} and velocity
  bias,'' \emph{IEEE Transactions on Intelligent Transportation Systems},
  vol.~PP, no.~PP, pp. 1--10, 2020.

\bibitem{nielsen2018ground}
J.~Nielsen and R.~Beard, ``Ground target tracking using a monocular camera and
  imu in a nonlinear observer slam framework,'' in \emph{2018 Annual American
  Control Conference (ACC)}.\hskip 1em plus 0.5em minus 0.4em\relax IEEE, 2018,
  pp. 6457--6462.

\bibitem{hashim2020SE3Stochastic}
H.~A. Hashim and F.~L. Lewis, ``Nonlinear stochastic estimators on the special
  euclidean group {SE}(3) using uncertain {IMU} and vision measurements,''
  \emph{IEEE Transactions on Systems, Man, and Cybernetics: Systems}, vol.~PP,
  no.~PP, pp. 1--14, 2020.

\bibitem{hashim2019SO3Wiley}
H.~A. Hashim, ``Systematic convergence of nonlinear stochastic estimators on
  the special orthogonal group {SO}(3),'' \emph{International Journal of Robust
  and Nonlinear Control}, vol.~30, no.~10, pp. 3848--3870, 2020.

\bibitem{odry2018kalman}
A.~Odry, R.~Fuller, I.~J. Rudas, and P.~Odry, ``Kalman filter for mobile-robot
  attitude estimation: Novel optimized and adaptive solutions,''
  \emph{Mechanical systems and signal processing}, vol. 110, pp. 569--589,
  2018.

\bibitem{odry2021open}
A.~Odry, ``An open-source test environment for effective development of
  marg-based algorithms,'' \emph{Sensors}, vol.~21, no.~4, p. 1183, 2021.

\bibitem{eltoukhy2018joint}
A.~E. Eltoukhy, Z.~Wang, F.~T. Chan, and S.~H. Chung, ``Joint optimization
  using a leader--follower stackelberg game for coordinated configuration of
  stochastic operational aircraft maintenance routing and maintenance
  staffing,'' \emph{Computers \& Industrial Engineering}, vol. 125, pp. 46--68,
  2018.

\bibitem{eltoukhy2019robust}
A.~E. Eltoukhy, Z.~Wang, F.~T. Chan, S.~H. Chung, H.-L. Ma, and X.~Wang,
  ``Robust aircraft maintenance routing problem using a turn-around time
  reduction approach,'' \emph{IEEE Transactions on Systems, Man, and
  Cybernetics: Systems}, vol.~50, no.~12, pp. 4919--4932, 2019.

\bibitem{ito1984lectures}
K.~Ito and K.~M. Rao, \emph{Lectures on stochastic processes}.\hskip 1em plus
  0.5em minus 0.4em\relax Tata institute of fundamental research, 1984,
  vol.~24.

\bibitem{deng2001stabilization}
H.~Deng, M.~Krstic, and R.~J. Williams, ``Stabilization of stochastic nonlinear
  systems driven by noise of unknown covariance,'' \emph{IEEE Transactions on
  Automatic Control}, vol.~46, no.~8, pp. 1237--1253, 2001.

\bibitem{hashim2018SO3Stochastic}
H.~A. Hashim, L.~J. Brown, and K.~McIsaac, ``Nonlinear stochastic attitude
  filters on the special orthogonal group 3: {I}to and {S}tratonovich,''
  \emph{IEEE Transactions on Systems, Man, and Cybernetics: Systems}, vol.~49,
  no.~9, pp. 1853--1865, 2019.

\end{thebibliography}
% name your BibTeX data base
\end{document}